\documentclass[journal=jacsat,manuscript=article,layout=onecolumn]{achemso}

\usepackage{chemformula} % Formula subscripts using \ch{}
\usepackage[T1]{fontenc} % Use modern font 
\usepackage{amsmath}
\usepackage{amssymb}
\usepackage{subfigure}
	\author{Arjun Varma Ramasimha Varma}
	\affiliation{Department of Metallurgical 
	Engineering and Materials Science, Indian 
	Institute of Technology Bombay, Mumbai 
	400076, India%\textbackslash\textbackslash
	}%Lines break automatically or can be forced 
	\author{Shilpa Paul}%
	\affiliation{%
		Department of Metallurgical Engineering 
		and Materials Science, Indian Institute 
		of Technology Bombay, Mumbai 400076, 
		India%\textbackslash\textbackslash
	}%
	
	\author{Anup Itale}%
	\affiliation{%
		Department of Metallurgical Engineering 
		and Materials Science, Indian Institute 
		of Technology Bombay, Mumbai 400076, 
		India%\textbackslash\textbackslash
	}%
	\author{Pranav Pable}%
	\affiliation{%
		Department of Metallurgical Engineering 
		and Materials Science, Indian Institute 
		of Technology Bombay, Mumbai 400076, 
		India%\textbackslash\textbackslash
	}%

	\author{Radhika Tibrewala}%
	\affiliation{%
	Department of Metallurgical Engineering and 
	Materials Science, Indian Institute of 
	Technology Bombay, Mumbai 400076, 
	India%\textbackslash\textbackslash
	}%

	\author{Samruddhi Dodal}%
	\affiliation{%
		Department of Metallurgical Engineering 
		and Materials Science, Indian Institute 
		of Technology Bombay, Mumbai 400076, 
		India%\textbackslash\textbackslash
	}%
	
	\author{Harshal Yerunkar}%
	\affiliation{%
		Department of Metallurgical Engineering 
		and Materials Science, Indian Institute 
		of Technology Bombay, Mumbai 400076, 
		India%\textbackslash\textbackslash
	}%

	\author{Saurav Bhaumik}%
	\affiliation{%
		Department of Mathematics, Indian 
		Institute of Technology Bombay, Mumbai 
		400076, India 
	}%

	\author{Vaishali Shah}%
	\affiliation{%
		Department of Scientific Computing, 
		Modeling and Simulation, Savitribai Phule 
		Pune University, Pune 411007, 
		India%\textbackslash\textbackslash
	}%

	\author{Mogadalai Pandurangan Gururajan}%
	\affiliation{%
		Department of Metallurgical Engineering 
		and Materials Science, Indian Institute 
		of Technology Bombay, Mumbai 400076, 
		India%\textbackslash\textbackslash
	}%

	\author{Tiramkudlu R. S. Prasanna}%
	\email{prasanna@iitb.ac.in}
	\affiliation{%
		Department of Metallurgical Engineering 
		and Materials Science, Indian Institute 
		of Technology Bombay, Mumbai 400076, 
		India%\textbackslash\textbackslash
	}%
	\date{\today}
\title{Electron-phonon interaction contribution 
to the total energy of group IV semiconductor 
polymorphs: evaluation and implications}
\abbreviations{}
\keywords{}
\begin{document}
%\twocolumn[
%\begin{@twocolumnfalse}
\begin{abstract}
 In density functional theory (DFT) based total energy studies, the van der 
 Waals (vdW) and zero-point vibrational energy (ZPVE) correction terms are 
 included to obtain energy differences between polymorphs. We propose and 
 compute a new  correction term to the total energy, due to electron-phonon 
 interactions (EPI). We rely on Allen's general formalism, which goes beyond 
 the Quasi-Harmonic Approximation (QHA), to include the free energy 
 contributions due to quasiparticle interactions. We show that, for 
 semiconductors and insulators, the EPI contributions to the free energies of 
 electrons and phonons are the corresponding zero-point energy contributions. 
 Using an approximate version of Allen's formalism in combination with the 
 Allen-Heine theory for EPI corrections, we calculate the zero-point EPI 
 corrections to the total energy for cubic and hexagonal  polytypes of Carbon, 
 Silicon and Silicon Carbide. The EPI corrections alter the energy differences 
 between polytypes. In SiC polytypes, the EPI correction term is more sensitive 
 to crystal structure than the vdW and ZPVE terms and is thus essential in 
 determining their energy differences. It clearly establishes that the cubic 
 SiC-3C is metastable and hexagonal SiC-4H is the stable polytype. Our results 
 are consistent with the experimental results of Kleykamp. Our study enables 
 the inclusion of EPI corrections as a separate term in the free energy 
 expression. This opens the way to go beyond the QHA by including the 
 contribution of EPI on all thermodynamic properties.  % 
\end{abstract}
%\end{@twocolumnfalse}

	\section{Introduction}
	
	Group IV semiconductors, especially C, Si 
	and SiC, are of immense 
	scientific and technological importance. 
	Currently, there is enormous interest in 
	predicting new allotropes of C and Si  
	because different crystal structures exhibit 
	differing physical properties and provide a 
	landscape to design materials with specific 
	properties 
	\cite{takagi2017,Mujica2015,he2018,dmitrienko2020,haberl2016}.
	 In the case of C, this is evident from 
	the 522 allotropes listed in the Samara 
	Carbon allotrope database \cite{Proserpio}. 
	In Si, predicting metastable crystal 
	structures with direct band-gaps are of particular
	interest 
	\cite{lee2014}. SiC polytypes are among the 
	most important materials for 
	structural applications \cite{riley2009}. SiC 
	also has promising potential in future 
	high-voltage and low-loss power devices 
	\cite{kimoto2015l}.
	
	Given their importance, several studies have 
	been performed on C, Si and SiC 
	polytypes.  The accurate determination of 
	the energy differences is essential in the 
	study of these polytypes. The main SiC polytypes are the cubic zincblende 
	SiC-3C, the hexagonal wurtzite SiC-2H and the hexagonal SiC-4H and SiC-6H.
 Despite several experimental and computational studies, the stable structure 
 in 
	SiC remains controversial because of the marginal energy differences 
	between the cubic (SiC-3C) and hexagonal SiC polytypes (SiC-4H and SiC-6H). 
	We first summarize the results from experimental studies and later from 
	computational studies.
		
	Greenberg et al. \cite{Greenberg} report thermodynamic data at 298 K for 
	hexagonal $\alpha$-SiC  and cubic $\beta$-SiC (SiC-3C) using calorimetric 
	studies. Using the quadrature rule for error propagation, the free energy 
	of transformation can be calculated from their results to be $\Delta_{tr} 
	G^{0, \alpha \rightarrow \beta}_{298}$ = -1.13 $\pm$ 2.6 kJ/mol (-11.70 
	$\pm$ 26.94 meV/formula unit, f.u.) and the 
	enthalpy of transformation is  $\Delta_{tr} H^{0, \alpha \rightarrow 
	\beta}_{298}$ = -1.09 $\pm$ 2.6 kJ/mol (-11.29 $\pm$ 26.94 meV/f.u.) using 
	the relation 1 kJ/mol = 10.36 meV/f.u. (Henceforth, in all energy data 
	given in units 
	of meV/f.u., f.u. refers to one unit of SiC, i.e. one atom of Si and C). 
	From the JANAF-Tables, \cite{JANAF1985janaf, Kleykamp1998gibbs}  the 
	enthalpy of transformation is 
	obtained as $\Delta_{tr} H^{0, \alpha \rightarrow \beta}_{298}$ = -1.7 
	$\pm$ 8.9 kJ/mol (-17.61 $\pm$ 92.20 meV/f.u.). (Since Greenberg et 
	al. \cite{Greenberg} and the 
	JANAF-Tables,  \cite{JANAF1985janaf} have not specified the polytype (2H, 
	4H, or 6H) for the hexagonal $\alpha$-SiC, we cannot be more 
	specific about the particular hexagonal SiC polytype.) From these results, 
	we may conclude that the cubic $\beta$-SiC (SiC-3C) has lower enthalpy. 
	However, the large margin of error suggests that the experimental results 
	from calorimetric studies are not conclusive but are indicative.
	
	In later studies, Kleykamp \cite{Kleykamp1998gibbs} has reported the 
	results of galvanic cell measurements and found that hexagonal $\alpha$-SiC 
	is stable. (Kleykamp \cite{Kleykamp1998gibbs} has reported that the 
	hexagonal $\alpha$-SiC used in his study was mainly SiC-6H.) In his study, 
	the stable structure was obtained directly from emf measurements in 
	galvanic cells. In the temperature range of 1100 K - 1300 K, the cell 
	arrangement with SiC-3C in the negative electrode and SiC-6H in the 
	positive electrode gave a positive emf of 20 $\pm$ 5 mV (indicating a 
	spontaneous process) leading to a Gibbs free energy of transformation 
	$\Delta_{tr} G^{3C \rightarrow 6H}$ = -8 $\pm$ 2 kJ/mol (82.88 $\pm$ 20.72 
	meV/f.u.). Unlike in most studies, in his study the stability of SiC-6H is 
	outside the margin of 
	error. Kleykamp \cite{Kleykamp1998gibbs} has reported that SiC-6H is also 
	stable at room temperature. For comparison, if the cubic SiC-3C were the 
	stable polytype, there would be no positive emf in the above cell 
	arrangement. Because of the contrasting nature of the results in 
	electrochemical studies, such experiments give clear indication of the 
	stable polytype. This highlights the importance of Kleykamp's study 
	\cite{Kleykamp1998gibbs}. A review of various methods to 
	obtain thermodynamic data for ceramic systems considers  
Kleykamp’s study as an elegant method to obtain small free energy differences 
and states that this 
was the first time the free energy difference between the two SiC phases was 
directly measured \cite{Jacobson2005307}.

	Due to these disagreements in the experimental results, a recent view is 
	that there is no general agreement on the stable SiC polytype 
	\cite{Drue2015phase}. However, based on the above discussion, it is clear 
	that the available 
	experimental 
	data in favor of SiC-6H as the stable polytype is much stronger. If 
	the results of Kleykamp 
	\cite{Kleykamp1998gibbs} are confirmed by independent electrochemical 
	studies, they will clearly establish the stability of SiC-6H under ambient 
	conditions. 	
	
	Computational studies using various DFT codes have consistently 
	shown that the hexagonal SiC-6H and SiC-4H 
	are marginally stable compared to cubic SiC-3C by a few 
	meV/f.u. \cite{Park1994, Kackell1994, Kawanishi2016, 
	Scalise2019, Heine1991, Rutter1997, Heine1992computational, Ramakers}. 
	These studies also show that the SiC-4H and SiC-6H polytypes are almost 
	degenerate. (For this reason, the \textit{ab initio} results for either 
	SiC-4H or 
	SiC-6H have been compared with experimental results in literature.)	
Kawanishi et al. \cite{Kawanishi2016} and Scalise et 
	al. \cite{Scalise2019} included the van der Waals (vdW) interaction 
	through the DFT-D2 approximation and found that SiC-3C is the stable 
	polytype. However, Ramakers et al. 
	\cite{Ramakers} 
	have recently studied the stability using ten different vdW 
	approximations including DFT-D2, DFT-D3, DFT-D3(BJ) and the advanced 
	many-body dispersion (MBD) approximations. They conclude that vdW 
	approximations do not have any significant effect on the relative 
	stabilities obtained from DFT 
	studies.  They also conclude that SiC-3C is metastable at all temperatures 
	after vibrational free energy contributions are included, in agreement with 
	the 
	results of Heine et al. \cite{Heine1991, 
	Heine1992computational,Rutter1997}. We note that these results are 
	consistent with the experimental results of Kleykamp 
	\cite{Kleykamp1998gibbs} discussed above, though the energy differences 
	between polytypes are much smaller in DFT studies.

	 Recently, for materials where polymorphs differ 
	marginally in energy (SiC, BN, B, 
	Fe\textsubscript{2}P etc.), the stable 
	polymorph has been determined by including 
	the vdW and the zero-point vibrational energy (ZPVE) 
	corrections. Including these corrections frequently alters the 
	polymorph stability order \cite{Kawanishi2016, Scalise2019, cazorla2019, 
	Nikaido2022, van2007, bhat2018}.
	
	However, these studies do not consider the contributions from 
	electron-phonon interactions (EPI). Electron-phonon interactions have been 
	well studied for their role in  several electronic and optical properties 
	\cite{Giustino2017,Ponce2014a}. In semiconductors and insulators, the 
	experimental observation of the temperature dependence of band gaps is 
	explained by EPI which leads to temperature dependent eigenenergies given 
	by $E_{n\textbf{k}}(T) = \epsilon_{n\textbf{k}} + \Delta 
	\epsilon_{n\textbf{k}}(T)$ where $\epsilon_{n\textbf{k}}$ is the static 
	lattice eigenenergy for wave vector $\mathbf{k}$ and band $n$ and the 
	second term is the EPI correction to the eigenenergy 
	\cite{Giustino2017,Ponce2014a}. Due to the presence of zero-point 
	vibrations, EPI alters the eigenenergies at 0 K as well. Because all the 
	eigenenergies are altered, it follows that the total energy will also be 
	altered	when EPI contributions are included. Hence, the EPI 
	contributions to the total energy and other thermodynamic properties must 
	be studied.
	
	In this paper, we propose and compute, for the first time, EPI corrections 
	to the total energy and free energy. In order to do so, we briefly describe 
	the the Quasi-Harmonic Approximation (QHA). We then discuss Allen's general 
	formalism that goes beyond QHA to include the contributions from 
	quasiparticle interactions (electron-phonon interactions, phonon-phonon 
	interactions leading to anharmonicity) to the  free energy expression 
	\cite{Allen2020, Allen2022erratum, Allen2022}. We then show that, in 
	semiconductors and 	insulators, the EPI contributions to the total 
	energy and free energy can 
	be obtained by combining Allen's expression for the free energy with the 
	Allen-Heine theory. We use this approach to calculate the EPI correction to 
	the total energy for C, Si and SiC polytypes.
	
	Our results show that including the EPI corrections alters the energy 
	differences between 
	polytypes in C, Si and SiC. For SiC polytypes, the EPI term is 
	more important than the ZPVE and the vdW correction terms, in 
	determining relative 
	stability. Inclusion of the EPI contribution clearly establishes the 
	metastability of the cubic SiC-3C polytype in 
	\textit{ab initio} studies. Our approach 
	enables the inclusion of EPI 
	corrections as a separate term in the free 
	energy expression. This opens the way to go beyond QHA by including the 
	contribution of EPI on all thermodynamic properties.

\section{Brief description of QHA}
In the QHA, the Helmholtz free energy is given by \cite{nath2016, Togo2015}
\begin{equation}
	F(V,T) = E_{0K}(V) + F_{el}^{QH}(V,T) + F_{vib}(V,T)
	\label{Free En-QHA}
\end{equation}

The $E_{0K}(V)$ term is the total energy at 0 K for a static lattice that is 
usually calculated from density functional theory (DFT), $F_{el}^{QH}(V,T)$ is 
the contribution to the free energy due to electronic 
excitations. The first two terms taken together give the independent 
particle electronic free energy. The third term, $F_{vib}(V,T)$, is the 
vibrational free energy contribution from harmonic (non-interacting) phonons. 
In Eq. \ref{Free En-QHA}, the $F_{el}^{QH}(V,T)$ term is calculated from the 
relation 
\cite{nath2016}
\begin{equation}
	F_{el}^{QH}(V,T) = E_{el}^{QH}(V,T) - TS_{el}^{QH}(V,T)
	\label{Free En-elec-QHA}
\end{equation}

Both the RHS terms are obtained from the electronic density of states (eDOS) 
at 0 K as \cite{nath2016}
\begin{equation}
	 E_{el}^{QH}(V,T) = \int n(\epsilon) f \epsilon d\epsilon - 
	\int_{0}^{\epsilon_F} n(\epsilon) \epsilon d\epsilon
	\label{Delta-E-elec}
\end{equation}

and 
\begin{equation}
	S_{el}^{QH}(V,T) = -k_B \int n(\epsilon) \left[ f ln f + 
	(1-f)ln(1-f)\right] d\epsilon
	\label{Delta-S-elec}
\end{equation}

where $f$ is the Fermi-Dirac distribution.

Allen \cite{Allen2020} has shown that the free energy expression in QHA can be 
derived in two different ways. The starting point is Eq.1 of Allen 
\cite{Allen2020}, that is the general expression for the eigenenergy given by
\begin{equation}
	\epsilon^{QP}_{K}(V,T) = \epsilon_K(V_0) + 
	\Delta \epsilon_K^{QH}(V) + \Delta 
	\epsilon_K^{QP}(V,T)
	\label{eqn-eigenen}
\end{equation}

where K represents (\textbf{k}, n), the wave-vector and band index 
respectively. The first term on the right-hand side is the eigenenergy for the 
equilibrium static lattice parameters. The second term is the change 
in eigenenergy due to the change in equilibrium volume (or lattice parameters)  
at finite temperatures. The sum of the first two terms is represented as 
$\epsilon_K(V) = \epsilon_K(V_0) + \Delta \epsilon_K^{QH}(V)$ and 
depends implicitly on temperature through the temperature 
dependence of the equilibrium volume, V(T)  \cite{Allen2020}. Thus, 
$\epsilon_K(V)$ is the 
eigenenergy of non-interacting electrons that is used in QHA.

The third term is the contribution from quasiparticle interactions (e.g. EPI, 
phonon-phonon interactions). This term is ignored in the QHA which is based on 
the assumption of non-interacting particles.

For non-interacting particles, the free energy can be obtained from two 
different approaches \cite{Allen2020}, i) from statistical mechanics and ii) by integrating the expression for the entropy.

 In the statistical mechanics approach, the free energy of both electrons and 
 phonons can be obtained  \cite{Allen2020} for the case of non-interacting 
 particles. We request the reader to refer to  Allen \cite{Allen2020} whose Eq. 
 11 gives the electronic free energy and Eq. 12 gives the vibrational free 
 energy. Taken together, they lead to the standard expression used in QHA, 
Eq. \ref{Free En-QHA} above.

Alternately, the free energy can also be obtained by integrating the expression 
for entropy \cite{Allen2020}. In this approach, the free energy of electrons 
and phonons are obtained using the entropy formulas for non-interacting 
electrons and phonons (Eq. 23 and Eq. 24 respectively of Allen 
\cite{Allen2020}) in the free energy expression, Eq. 25 of Allen (also given 
below).

The two approaches to obtain the free energy are equivalent for non-interacting 
particles \cite{Allen2020}.   

\section{EPI contributions to the free and total energy}

Allen \cite{Allen2020, Allen2022erratum} has described a method to go beyond 
the QHA and incorporate the renormalization of eigenenergies due to 
quasiparticle interactions (electron-phonon interactions, phonon-phonon 
interactions leading to anharmonicity) in the free energy expression.

For interacting quasiparticles, the eigenenergy, $\epsilon^{QP}_{K}(V,T)$, is 
obtained by taking all the three terms in Eq. \ref{eqn-eigenen} and is 
temperature dependent.

 A limitation of the statistical mechanics based approach to obtain free energy 
 is that it is only valid for non-interacting electrons and phonons and not for 
 the case of interacting quasiparticles. However, the alternate approach based 
 on the entropy formula can be applied for the case of interacting particles, 
 provided the quasiparticles have well-defined energies (when the lifetime 
 broadening is not very large) \cite{Allen2020, Allen2022}. The Helmholtz free 
 energy (Eq. 25 of Allen \cite{Allen2020}) is obtained by integrating the 
 expression for the entropy and given as
\begin{equation}
	F^{QP}(V,T) = F^{QP}(V,0) - \int_{0}^{T} 
	S_{el}^{QP}(V,T^{\prime}) dT^{\prime}
	\label{Free energy-QP-Allen}
\end{equation}

where the entropy is given by \cite{Allen2020}
\begin{equation}
	S_{el}(V,T) = -k_B \sum_K \left[ f_K ln f_K + (1-f_K)ln(1-f_K)\right]
	\label{Entropy-Allen}
\end{equation}

and $f_K$ is the Fermi-Dirac distribution. For the case of interacting 
quasiparticles, the temperature dependent eigenenergy, Eq. \ref{eqn-eigenen}, 
must be used \cite{Allen2020}.

The corrected term $F^{QP}(V,0)$ in Eq. \ref{Free 
energy-QP-Allen} is given by \cite{Allen2022erratum}
\begin{equation}
	F^{QP}(V,0) = E^{QP}(V,0) = E_{0K}(V) + \Delta E_{EP}(V,0)
	\label{Free energy-QP-Allen - 0K}
\end{equation}

where $\Delta E_{EP}(V,0)$ is the zero-point EPI contribution to the total 
energy given by \cite{Allen2022erratum}
\begin{equation}
	\Delta E_{EP} (V,0) = \sum_k \left[\langle k 
|V^{(2)}|k\rangle + \sum_Q \frac{|\langle k 
|V^{(1)}|k+Q\rangle|^2}{\epsilon_k - 
\epsilon_{k+Q}}(1-f_{k+Q}) \right]\;f_k
	\label{EPI-zero-K-Allen}
\end{equation}

The term within the square bracket in the RHS of Eq. \ref{EPI-zero-K-Allen} is 
very similar (but not identical) to the original expression for the EPI 
correction to the eigenenergy ($\Delta \epsilon_{n\textbf{k}}$) \cite{Allen1976} and differs in the presence of the factor ($1-f_{k+Q}$) in the second term \cite{Allen2022erratum}.

For semiconductors and insulators, the band-gaps are large and the electron 
occupancies do not change with temperature. Thus, $f_K$ = 1 at all temperatures 
and hence, the entropy contribution, Eq. \ref{Entropy-Allen}, is zero. Thus, 
for semiconductors and insulators, the free energy of electrons when EPI is 
included reduces to  
\begin{equation}
	F^{QP}_{SC,I}(V,T) = E_{0K}(V) + \Delta E_{EP}(V,0) 
	\label{Free energy-QP-Final-SemiCond}
\end{equation}

For semiconductors and insulators, comparing Eq. \ref{Free 
energy-QP-Final-SemiCond} with the electronic free energy for 
non-interacting particles used in QHA (Eq. \ref{Free En-QHA}), the free energy 
contribution due to EPI is an additional term, $\Delta E_{EP}(V,0)$, the 
zero-point EPI contribution to the total energy. For semiconductors and 
insulators, the electron occupancies do not vary with temperature and hence, 
the temperature dependence of Eq. \ref{Free energy-QP-Final-SemiCond} is only 
due to the changes in equilibrium volume with temperature, V(T).

We next discuss the EPI contribution to the phonon energies. The EPI 
corrections to the phonon frequencies and energies consists of two parts, 
adiabatic and non-adiabatic components \cite{Giustino2017, Allen2022, 
Heid201312}. Of these, the adiabatic component is already included in a density 
functional perturbation theory (DFPT) calculation \cite{Giustino2017, 
Allen2020, Allen2022, Heid201312}. The non-adiabatic component 
can be obtained by using time-dependent DFT \cite{Heid201312, Saitta2008giant, 
Calandra2010adiabatic}. However, the non-adiabatic correction is important 
only in metals and narrow band-gap semiconductors \cite{Giustino2017, 
Calandra2010adiabatic}. It is negligible in insulators and large band-gap 
semiconductors where the band gap is much larger than the phonon energy 
\cite{Giustino2017, Calandra2010adiabatic}. Thus, for large band-gap 
semiconductors and insulators, only the adiabatic component of the EPI 
contribution to phonon energies must be considered and it is already included 
in a DFPT calculation \cite{Giustino2017, Allen2020, Allen2022, Heid201312}.

The phonon spectra and the vibrational free energy 
obtained from DFPT and other methods, e.g bond-charge model (BCM) and finite 
difference (FD), are reported to be similar in literature \cite{Karch1996, 
Li2019continuity}.Any method to calculate the phonon spectra and phonon free energies that gives identical values to the DFPT method implicitly includes the adiabatic phonon contribution \cite{Allen2022}. It follows that the 
vibrational free energy can be calculated from any of the above methods when 
EPI is included.

Allen and Hui \cite{Allen1980} have shown that the EPI contribution to the 
electronic and vibrational specific heats are equal to order $\langle u^2 
\rangle$ i.e., $\Delta C_{el}^{EP} = \Delta C_{ph}^{EP}$. For semiconductors 
and insulators, $\Delta C_{el}^{EP}$ is zero due to the 
$f(\epsilon)\left[1-f(\epsilon)\right]$ factor \cite{Allen1980} and hence, 
$\Delta C_{ph}^{EP}$ is also zero. Thus, the EPI contribution to the phonon 
free energy will be a 
zero-point correction, which is included in a DFPT calculation. 

In summary, for semiconductors and insulators, because $\Delta C_{el}^{EP} = 
\Delta C_{ph}^{EP} = 0$, the EPI contributions to the free energies of 
electrons and phonons are the corresponding zero-point energy contributions. 
The temperature dependence of the EPI contributions will only come from the 
changes in the volume, V(T) \cite{Allen2022}.

An important consequence for semiconductors and insulators is that any 
experimentally observed anomalous temperature dependence of the phonon spectra 
is most likely due to other quasiparticle interactions, e.g., phonon-phonon 
interactions leading to anharmonicity.

Another important implication follows for QHA studies where the free energy is 
given by Eq. \ref{Free En-QHA}, though sometimes the vdW contributions  are 
also included in the DFT total energy. The EPI contribution to the 
electronic free energy must be neglected in QHA. However, the 
vibrational free energy calculated using DFPT and other (FD, BCM etc.) methods 
include, explicitly and implicitly respectively, the adiabatic EPI contribution 
\cite{Allen2022}. To be consistent with the neglect of EPI contribution to the 
electronic free energy, the adiabatic EPI contribution to the phonon spectra 
and vibrational free energy must be subtracted in QHA studies. 

\section{EPI corrections to the total energy from the Allen-Heine theory}

	The Allen-Heine theory \cite{Allen1976, Allen1978, Allen1980, Allen1983, 
	Allen1994} is widely used to obtain the EPI corrections  to the 
	eigenenergies in semiconductors and insulators. In this theory,  
	electron-phonon interactions lead to 
	contributions from the Fan-Migdal (FM) 
	\cite{Fan1951,Migdal1958} and the 
	Debye-Waller (DW) \cite{Antoncik1955} terms 
	to the eigenenergies. The temperature 
	dependent eigenenergies are given by  
	$E_{n\textbf{k}}(T) = \epsilon_{n\textbf{k}} 
	+ \Delta \epsilon_{n\textbf{k}}(T)$ 
	where $\epsilon_{n\textbf{k}}$ is the static 
	lattice eigenenergy for wave vector 
	$\mathbf{k}$ and band $n$ and the EPI 
	correction is given by 
	$\Delta \epsilon_{n\textbf{k}}(T) = 
	\Delta^{FM}\epsilon_{n\textbf{k}}(T)+\Delta^{DW}\epsilon_{n\textbf{k}}(T)$.
	Due to the presence of zero-point vibrations, 
	the zero-point renormalization (ZPR) of 
	electron eigenenergies, $\Delta 
	\epsilon_{n\textbf{k}}(0)$, has finite 
	values. 
		 
	The Allen-Heine theory allows the calculation of EPI correction to any 
	 arbitrary eigenstate, $\Delta \epsilon_{n\textbf{k}}(T)$, including for 
	 the valence band maxima (VBM) and the  conduction band minima (CBM). For 
	 this  reason, it is the most widely used  \textit{ab initio} method in EPI 
	 studies of  band gaps and band structures  \cite{Marini2008, Giustino2010, 
	 Antonius2014, Ponce2014b, Ponce2015, Antonius2015,
	 Friedrich2015, Allen2018, Tutchton2018, Querales2019, Gonze2020, 
	 Cannuccia2020thermal}. We note that there are other recent methods to 
	 obtain the EPI contributions to band-gaps and band structures 
	 \cite{Patrick2014,Zacharias2016}.
	 
	The zero-point EPI contribution to the total energy, $\Delta E_{EP} (V,0)$ 
	is given by Eq. \ref{EPI-zero-K-Allen} where the second term has a 
	$1-f_{k+Q}$ factor that is not present in the original expression for 
	$\Delta \epsilon_{n\textbf{k}}(T)$ \cite{Allen2022erratum}. This implies 
	that the present methods used to calculate $\Delta 
	\epsilon_{n\textbf{k}}(T)$ cannot be used to obtain the zero-point EPI 
	contribution to the total energy,  $\Delta E_{EP} (V,0)$. The correct 
	evaluation of $\Delta E_{EP} (V,0)$ will be possible for the wider 
	computational community only after the computation of Eq. 
	\ref{EPI-zero-K-Allen} is included in software packages. 
  
	 Therefore, in the present paper, we calculate an approximation to $\Delta 
	 E_{EP} (V,0)$ by summing over the eigenenergies, $\Delta 
	 \epsilon_{n\textbf{k}}(T)$. In this case,  $\Delta E_{EP} (V,0)$ is given 
	 by

	\begin{equation}
	\Delta E_{EP} (V,0) \approx \Delta E^{ep}_{av}(V,0) = \sum^{occ}_K \Delta 
	\epsilon_{n\textbf{k}}(V,0) = 2 
	\sum_{n,\textbf{k}}^{occ,IBZ} w_\textbf{k}\Delta \epsilon_{n\textbf{k}} 
	(V,0) 
	\label{EPI-approx-Abinit} 
	\end{equation}

where $\Delta E_{EP} (V,0)$ is approximated by the zero-point EPI correction to 
the band-structure energy, $\Delta E^{ep}_{av}(V,0)$, $w_\textbf{k}$ is the 
weight of the \textbf{k}-point in the irreducible Brillouin Zone (IBZ) and 
$\Delta \epsilon_{n\textbf{k}} (V,0)$ is the zero-point EPI correction to the 
static lattice eigenenergy $\epsilon_{n\textbf{k}} (V)$.

The approximation, Eq. \ref{EPI-approx-Abinit} is likely to introduce some 
errors compared to Eq. \ref{EPI-zero-K-Allen} due to the neglect of the 
$1-f_{k+Q}$ factor. However, the main interest in relative stability studies is 
in the EPI contribution to the total energy differences between 
polytypes. In this case, the errors are expected to 
be much less due to the neglect of the $1-f_{k+Q}$ factor in all 
polytypes. Comparing Eq. \ref{EPI-zero-K-Allen} and Eq. 
\ref{EPI-approx-Abinit}, the contribution from the Debye-Waller term remains 
unchanged. The approximate expression, Eq. \ref{EPI-approx-Abinit} introduces 
errors only in the Fan-Migdal term, because some of the contributions that must 
be neglected are included. Thus, the error in using  Eq. 
\ref{EPI-approx-Abinit} is due to only a part of the Fan-Midgal term. Since 
this error is present in all polymorphs, the cancellation of errors for energy 
differences between polymorphs implies 
that the trends observed from using Eq. \ref{EPI-approx-Abinit} are likely to 
remain valid even when Eq. \ref{EPI-zero-K-Allen} is used. This is corroborated 
by the better match with experimental results when EPI corrections are included 
using Eq. \ref{EPI-approx-Abinit} compared to earlier DFT based studies 
\cite{Park1994, Kackell1994, Kawanishi2016, Scalise2019, Heine1991, Rutter1997, 
Heine1992computational, Ramakers} as seen below. Further justifications for 
specific systems are given later.

A perusal of the ZPR shifts, $\Delta \epsilon_{n\textbf{k}} (V,0)$, of all 
occupied states in the IBZ for C, Si and SiC polytypes (in this study) does not 
reveal any discernible trend. For each $\textbf{k}$-point, $\Delta 
\epsilon_{n\textbf{k}} (V,0)$ has positive and negative values for some 
occupied $n$-band. There is no $\textbf{k}$-point where  $\Delta 
\epsilon_{n\textbf{k}} (V,0)$ for the occupied $n$-bands are all positive or 
all negative. The zero-point EPI correction to the band-structure energy, 
$\Delta E^{ep}_{av}(V,0)$ is a weighted sum over several positive and 
negative values of $\Delta \epsilon_{n\textbf{k}} (V,0)$. Thus, presently, no 
general conclusions can be drawn about the effect of the $1-f_{k+Q}$ factor. 

It is also seen that $\Delta E^{ep}_{av}(V,0)$ for different polytypes are 
similar, but not equal. Further, the convergence behavior of $\Delta 
E^{ep}_{av}(V,0)$ is similar for different polytypes.  In this case it is 
justified to assume that the 
Debye-Waller contributions and also the Fan-Migdal contributions are also 
similar for different polytypes. That is, the Fan-Migdal contributions to the 
total energy from terms affected and unaffected by the $1-f_{k+Q}$ factor are 
also likely to be similar for different polytypes. Hence, as a first 
approximation, we can assume that the neglect of the $1-f_{k+Q}$ factor in Eq. 
\ref{EPI-approx-Abinit} is unlikely to significantly alter the energy 
differences between polytypes, due to cancellation of errors.  

Otherwise, it would imply that the terms affected by the $1-f_{k+Q}$ factor in  
Eq. \ref{EPI-zero-K-Allen} are the dominant terms in the EPI contribution to 
the energy differences between polymorphs even when  $\Delta E^{ep}_{av}(V,0)$ 
values are similar. At present, there is no basis for such a conclusion.

Therefore, in this study, we have calculated the EPI corrections to the 	 
total energy for the cubic and hexagonal polytypes of C, Si and SiC using Eq. 
\ref{EPI-approx-Abinit}.
	 
	 \section{Computational Details}
	 
	All calculations were performed using the ABINIT software package 
	\cite{Gonze2009, Gonze2016, Abinit} that has been used in several ZPR 
	band-gap studies \cite{Ponce2014b, Friedrich2015, Allen2018, Tutchton2018, 
	Querales2019, Gonze2020}.
	
	For SiC polytypes, the vdW approximations, DFT-D2 and DFT-D3(BJ), 
	\cite{Grimme2006,Grimme2011} were also applied to calculate the total 
	energies and lattice parameters. DFT-D\textit{n} methods were chosen 
	because of the advantage that the vdW and EPI corrections are both 
	incorporated by calculating the EPI corrections at the altered lattice 
	parameters \cite{Troeye2016, Tutchton2018}.
	
	We present results obtained using the ONCV pseudopotentials 	
	\cite{HamannONCV} with PBE exchange-correlation functional 
	\cite{PerdewPBE}. Similar results were obtained for other pseudopotentials. 
	(See Supplementary Information). The energy cutoffs used were 30 Ha (Si) 
	and 50 Ha (C and SiC). For hexagonal structures, the 
	Broyden-Fletcher-Goldfarb-Shanno (BFGS) algorithm was used for structural 
	optimization \cite{Gonze2009, Gonze2016}. An unshifted $8 \times 8 \times 
	8$ $\mathbf{k}$-point grid was used for cubic structures. For all hexagonal 
	structures (except for SiC-4H for which $9 \times 9\times3$ 
	$\mathbf{k}$-point grid was used), an unshifted $9 \times 9 \times 5$ 
	$\mathbf{k}$-point grid was used.
	
	The ABINIT module on temperature dependence of the electronic structure 
	that calculates EPI corrections using Allen-Heine theory was used 
	\cite{Abinit}. We provided the list of $\mathbf{k}$-points in the IBZ and 
	additional $\mathbf{k}$-points corresponding to CBM. The $\mathbf{q}$-point 
	grids were increased by 200-300 $\mathbf{q}$-points in IBZ. To accelerate 
	convergence, an imaginary smearing parameter, $i\delta$, of 100 meV and 50 
	meV was used \cite{Abinit,Ponce2015,Tutchton2018}. The optimized number of 
	bands used was 30 for SiC-4H and 22 for all other C, Si and SiC 
		polytypes. EPI corrections for SiC-6H 
		could not be calculated
		due to computational constraints.

	The adiabatic and non-adiabatic approximation must be used to obtain the 
	$\Delta \epsilon_{n\textbf{k}}(T)$ for IR-inactive and IR-active materials 
	respectively \cite{Ponce2015}. The adiabatic and non-adiabatic formulas to 
	calculate $\Delta \epsilon_{n\textbf{k}}(T)$ in ABINIT are given by Eqs. 
	15-17 of Ponce et al. \cite{Ponce2015}.

	For IR-active materials, the ZPR shift, $\Delta \epsilon_{n\textbf{k}}(T)$, 
	varies linearly with $1/N_q$, where $N_q^3$ is the total number of 
	$\mathbf{q}$-points in the BZ \cite{Ponce2015,Querales2019}. It follows 
	from the linearity property of Eq.  \ref{EPI-approx-Abinit} that $\Delta 
	E_{av}^{eq}(T)$ will also vary linearly with $1/N_q$ (Proof in 
	Supplementary Information). 

\section{Results and Discussion}
\subsection{EPI contributions in carbon and silicon polytypes}

	Table \ref{table1} gives the lattice parameters, band gaps and the 
	relative energy stability with and without EPI corrections at 0 K for the C 
	and Si polytypes. Our DFT lattice parameters and relative energies are 
	similar to those in literature \cite{Raffy2002,Mujica2015,Fan2018}.
	
	\begin{table}[h]
		\caption{\label{table1}%
			The lattice parameters, band gaps and energy differences of the 
			hexagonal polytypes with respect to the corresponding diamond 
			polytypes for carbon and silicon.}
		%\begin{ruledtabular}
		\begin{tabular}{p{2.0cm} c c c c c}
			\hline	
			\hline
			Material& a,c & \multicolumn{2}{c}{Band gaps}  & $\Delta E$ 
			(DFT) & $\Delta E$ 
			(DFT+EPI)\\
			& & Indirect & Direct & & \\
			& (Bohr)& (eV) & (eV) & (meV/atom) & (meV/atom) \\
			\hline
			C-dia & 6.75 & 4.18 & 5.61 & 0 & 0\\
			C-hex & 4.75, 7.90 & 3.40 & 5.01 & 24.3 & 
			67.9\\
			\hline
			Si-dia & 10.34 & 0.61 & 2.55 & 0 & 0\\   
			Si-hex & 7.28, 12.03 & 0.45 & 0.98 & 9.7 & 
			17.3\\
			\hline		
			\hline
		\end{tabular}
		%\end{ruledtabular}
	\end{table}
	
	For C-dia, the ZPR of VBM, CBM and the lowest CB at $\Gamma$-point are 142 
	meV, -151.4 meV and -272.5 meV respectively. For Si-dia they are 34.27 meV, 
	-21.24 meV and -8 meV respectively. These values are  similar to the 
	reported values \cite{Abinit}. Because, the ZPR for other eigenstates in 
	the IBZ were obtained in the same calculation, it follows that they are 
	reliable.

\begin{figure}[h]
		\includegraphics[width=7cm]{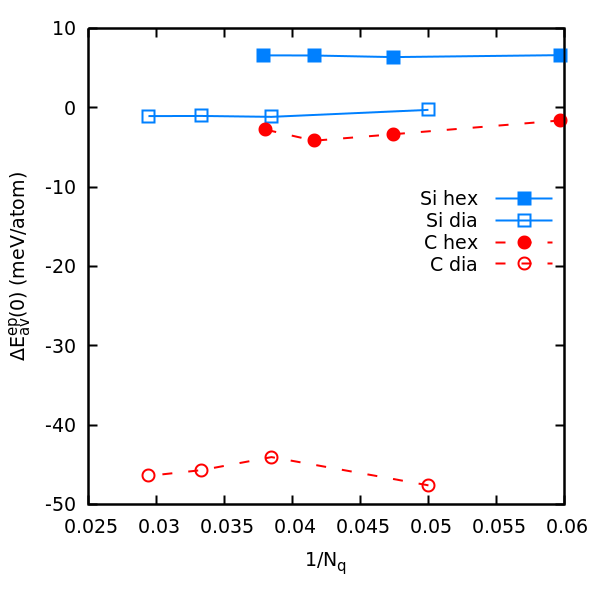}
		\caption{\label{fig1} Convergence of EPI correction to total energy at 0 K
		with $\mathbf{q}$-point grid density (adiabatic approximation) for a 
		smearing parameter of 100 meV for Carbon and Silicon polytypes.}
	\end{figure}

	Figure \ref{fig1} shows the convergence behavior of the EPI correction to 
	the total energy, $\Delta E^{ep}_{av}(0)$, with $1/N_q$ for $i\delta$ = 100 
	meV. A strong EPI correction is seen for carbon polytypes whereas a weak 
	EPI correction is seen for silicon polytypes. This is consistent with the 
	weaker electron-phonon interaction in silicon compared to carbon 
	\cite{Giustino2010,Antonius2014,Ponce2014b,Ponce2015,Antonius2015,Patrick2014,Zacharias2016,Friedrich2015,Allen2018,Tutchton2018}.
	 Figure \ref{fig1} also shows a significant crystal structure dependence of 
	the EPI correction to the total energy, especially in carbon polytypes 
	where the electron-phonon interaction is strong.
		
	The negative value of $\Delta E^{ep}_{av}(0)$ for C-dia implies that, 
	averaged over the BZ, the FM term is greater than the DW term. In contrast 
	to C-dia, a small value of $\Delta E^{ep}_{av}(0)$ is obtained for C-hex. 
	The ZPR of VBM, CBM and the lowest CB at $\Gamma$-point for C-hex are 
	117 meV, -172.6 meV and -313.9 meV respectively. They are comparable to 
	those for C-dia indicating a strong electron-phonon interaction in C-hex. 
	Thus, the small value of $\Delta E^{ep}_{av}(0)$ in C-hex indicates a near 
	balance between the FM and DW terms when averaged over the BZ.

	Figure \ref{fig1} shows	that the $\Delta E^{ep}_{av}(0)$ values for the 
	silicon polytypes are similar. Hence, their difference is unlikely to be 
	significantly affected when the correct expression, Eq. 
	\ref{EPI-zero-K-Allen}, that includes the $1-f_{k+Q}$ factor is used. 
	However, Figure \ref{fig1} shows that the $\Delta E^{ep}_{av}(0)$ values 
	for the carbon polytypes are not similar. It is a small negative value for 
	C-hex and relatively large negative for C-dia. One possibility is that the 
	large difference is due to the stronger EPI interaction in carbon. However, 
	another possibility is that using the correct expression, Eq. 
	\ref{EPI-zero-K-Allen}, that includes the $1-f_{k+Q}$ factor, could lead to 
	significant changes in the energy differences. Hence, the results for 
	carbon polytypes should be considered to be less reliable than the results 
	for silicon polytypes.

	Table \ref{table1} shows that after including EPI corrections the C-dia 
	structure is more stable than the C-hex structure by $\approx$ 68 meV/atom 
	compared to $\approx$ 24 meV/atom from DFT studies 
	\cite{Raffy2002,Mujica2015,Fan2018}. The Si-dia stability also increases 
	from $\approx$ 10 meV/atom to $\approx$ 17 meV/atom.  It clearly follows 
	that including EPI corrections in \textit{ab initio} studies of relative 
	stability of polymorphs is essential.

	\subsection{EPI contributions in SiC polytypes}
	
	Table \ref{table2} shows the lattice 
	parameters, ZPR of the VBM/CBM and the energy stabilities 
	relative to the SiC-3C polytype. The 
	indirect/direct band gaps (at the 
	$\Gamma$-point) are 
	1.41 eV/6.13 eV (SiC-3C), 2.32 eV/4.72 eV 
	(SiC-2H), 2.26 eV/5.01 eV (SiC-4H) and 2.07 
	eV/5.10 eV (SiC-6H). These results are very 
	similar to literature values 
	\cite{Park1994,Kackell1994,Kawanishi2016,Scalise2019}.
	
	\begin{table}[h]
		\caption{\label{table2}%
			Lattice parameters, ZPR and energy stability 
			of SiC polytypes relative
			to SiC-3C for DFT, DFT-D2 and DFT-D3(BJ) 
			calculations.}
		%\begin{ruledtabular}
		%\begin{tabular}{p{2.7cm}p{2.9cm}p{2.9cm}p{2.9cm}}
		\begin{tabular}{p{2.7cm} c c c}
			Polytype & a, c& ZPR 
			VBM/CBM &$\Delta E$ \\
			&(Bohr) &(meV) 
			& (meV/f.u.) \\
			\hline
			\textbf{3C-SiC} & {} & {} & {} \\
			DFT & 8.28 & 93.6/-56.3 & 0  \\
			DFT-D2 & 8.23 & - & 0 \\
			DFT-3(BJ) & 8.21 & 95.4/-55.6 & 0 \\
			%
			%\textbf{6H-SiC} & {} & {} & {} \\
			%
			%DFT & 5.85, 28.70 & -& -2.22  \\
			%DFT-D2 & 5.82, 28.56 & -& 0.72 \\
			%DFT-D3(BJ) & 5.80, 28.48 & -& -2.02 \\
			%
			\textbf{4H-SiC} & {} & {} & {} \\
			DFT & 5.85, 19.14 & 84.7/-61.3 & -2.15  \\
			DFT-D2 & 5.81, 19.06 & -& 2.31 \\
			DFT-D3(BJ) & 5.80, 18.99 & 86.3/-60.8 & -1.81 
			\\
			\textbf{2H-SiC} & {} & {} & {} \\
			DFT & 5.84, 9.59 & 78.6/-88.9 & 5.07  \\
			DFT-D2 & 5.81, 9.56 & -& 14.2 \\
			DFT-D3(BJ) & 5.80, 9.52 & 80.0/-86.5& 6.32 \\
		\end{tabular}
		%\end{ruledtabular}
	\end{table}
	
	In our DFT results, SiC-4H is more stable 
	than SiC-3C, similar to previous studies 
	\cite{Park1994,Kawanishi2016,Scalise2019,
	Kackell1994}. Including the DFT-D2 
	approximation makes SiC-3C to be the stable 
	polytype, similar to recent studies 
	\cite{Kawanishi2016,Scalise2019}. 
	However, including the DFT-D3(BJ) 
	approximation retains the DFT stability 
	order where SiC-3C is metastable, consistent with the results of Ramakers 
	et al. \cite{Ramakers}. As discussed
	earlier, Ramakers et al. \cite{Ramakers} 
	have considered ten different vdW approximations and concluded that 
	the DFT stability order should be retained.
				 
	 In Table \ref{table2}, the ZPR of VBM/CBM for SiC-3C {(obtained using 
	 parameters similar to those used in other studies~\cite{Tutchton2018})} is 
	 	comparable to reported values \cite{Gonze2020, Cannuccia2020thermal}.

\begin{figure}[h!]
		\subfigure[]{\includegraphics[width=8cm]{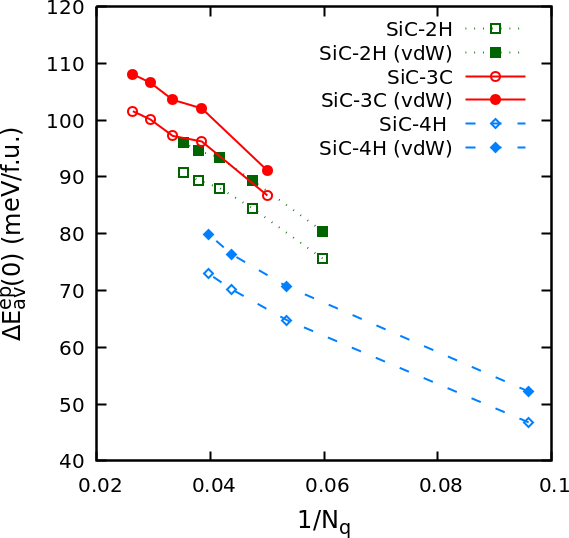}}
		\subfigure[]{\includegraphics[width=8cm]{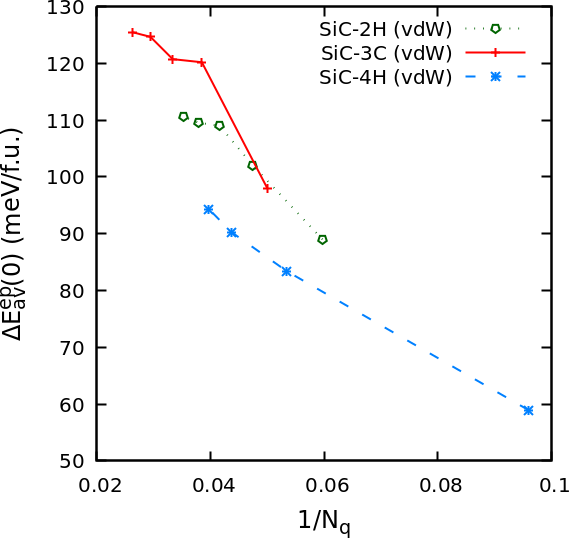}}
		\caption{\label{fig2} Convergence of EPI 
		correction to total energy at 0 K with 
		$\mathbf{q}$-point grid 
		density (non-adiabatic approximation) for 
		smearing parameters: a) 100 meV and b) 50 
		meV, for SiC polytypes for DFT 
		and DFT-D3(BJ) lattice parameters. }
	\end{figure}

Figure \ref{fig2} shows the convergence behavior of the EPI correction to the 
total energy, $\Delta E^{ep}_{av}(0)$. The $\Delta E^{ep}_{av}(0)$ for 
SiC-4H (50\% hexagonality)  \cite{Kawanishi2016,Scalise2019} does not lie 
between the values for SiC-3C (0\% hexagonality) and SiC-2H (100\% 
hexagonality). A similar trend is also seen for DFT total energies (Table 
\ref{table2} and Ref. \cite{Park1994,Kackell1994,Kawanishi2016,Scalise2019}). 
		 
 	Figure \ref{fig2} shows that $\Delta E^{ep}_{av}(0)$ varies substantially 
 	among SiC polytypes compared to the marginal DFT energy differences. For 
 	example, the EPI contribution to the total energy for SiC-3C and SiC-4H 
 	differ by $>$ 20 meV/f.u. This difference is much higher than the marginal 
 	(1-2 meV/f.u.) difference in the DFT total energy. 

		Figure \ref{fig2} shows that $\Delta E^{ep}_{av}(0)$ has similar  
	convergence behaviour for the SiC polytypes. This suggests that the DW and 
	FM terms have similar values in SiC-polytypes. In this regard, in their 
	Supplementary Table II, Miglio et al. 	\cite{Gonze2020} have compared the 
	calculated ZPR of the SiC-3C band gap with the experimental data for the 
	SiC-15R (40\% hexagonality) with the justification that the two SiC 
	polytypes have sp$^3$-bonding and differ only in the stacking sequence of 
	the tetrahedra \cite{Kawanishi2016, Ramakers}. Underlying the above 
	comparison is the implicit assumption that the ZPR (DW and FM terms) behave 
	similarly in SiC polytypes. The assumption of Miglio et al. 
	\cite{Gonze2020} that the structural similarity 	of SiC polytypes lead 
	to similar EPI behaviour is also seen in Fig. \ref{fig2} above. It follows 
	that the FM terms affected and unaffected by the $1-f_{k+Q}$ factor also 
	behave similarly in the SiC polytypes; which is our assumption.  In 
	structurally similar polytypes, the errors due to the neglect of the 
	$1-f_{k+Q}$ factor are likely to be similar, leading to small energy 
	differences among them due to cancellation 
	of errors. This provides additional justification for our assumption that 
	neglect of the $1-f_{k+Q}$ factor will not have a dominant effect on energy 
	differences. Thus, these are strong reasons to assume that using the 
	correct Eq. \ref{EPI-zero-K-Allen} is likely to lead to similar trends for 
	energy differences as obtained from the approximate Eq. 
	\ref{EPI-approx-Abinit}.
	
	Table \ref{table3} shows the stability of 
	SiC-polytypes obtained by combining the EPI 
	corrections to total energy in Figure 
	\ref{fig2} with the relative stability data 
	in Table \ref{table2}. 
	For SiC-3C and SiC-2H we consider $\Delta 
	E^{ep}_{av}(0)$ for the smallest value of 
	$1/N_q$. However, for SiC-4H, the $\Delta 
	E^{ep}_{av}(0)$ varies by $\lesssim$ 3 
	meV/f.u. for the smallest two $1/N_q$ values. 
	Therefore, we report two values for SiC-4H in 
	Table 
	\ref{table3}; the difference in $\Delta 
	E^{ep}_{av}(0)$ a) between the smallest 
	values of $1/N_q$ for SiC-3C and SiC-4H and 
	b) at approximately the same  $1/N_q$ 
	$\approx$ 0.039. 
	The actual stability value for SiC-4H is 
	likely to be between these values.

	\begin{table}[h]
		\caption{\label{table3}%
				Relative stability (meV/f.u.) of SiC polytypes for smearing 
				parameters of 100 meV and 50 meV calculated for the smallest 
				$1/N_q$ used. The second value in column-3 corresponds to the 
				relative stability at $1/N_q$ $\approx 0.039$ for both SiC 
				polytypes. The D2 
				values are estimated from DFT and D3(BJ) values as its lattice 
				parameters are in-between their lattice parameters.}		
		%\begin{ruledtabular}
			%\begin{tabular}{p{3.3cm}p{2.3cm}p{2.3cm}p{2.3cm}}
			\begin{tabular}{p{3.0 cm} c c c}
		    Polytype$\rightarrow$ \linebreak 
		    Stability $\downarrow$ & SiC-3C & 
			    SiC-4H & SiC-2H\\
				\hline
				% DFT             & 0  & -2.1 & 
				%5.1\\
				% DFT-D3(BJ)      & 0 & -1.8 & 
				%6.3\\
				 DFT + EPI \newline (100meV)      
				 & 0 & 
				 -30.8/-25.5 & -5.7\\ \\
				 DFT-D2+ \newline EPI(100 meV)  
				 & 0 & 
				 -25.6/-19.9 &  +2.8\\ \\
				 DFT-D3(BJ)+ \newline EPI (100meV) 
				 & 0 & 
				 -30.0/-24.0 & -5.6\\ \\
				 DFT-D3(BJ) + \newline EPI (50meV) 
				 & 0 & 
				 -32.9/-27.6 & -8.5
			\end{tabular}
		%\end{ruledtabular}
	\end{table}

	Table \ref{table3} shows the importance of EPI corrections. SiC-4H is the 
	stable polytype with much greater relative stability ($\geq$ 20 meV/f.u), 
	irrespective of the DFT-D approximation, compared to the marginal 
	stability ($\approx$ 2 meV/f.u.) under DFT and DFT+vdW conditions. For 
	SiC-2H, the relative stability depends on the DFT-D approximation used, 
	though it is more stable in the widely used D3(BJ) approximation 
	\cite{Voorhis,cazorla2019}. However, the stability is $<$ 10 meV/f.u., 
	indicating that it is within the range of dispersion approximation errors 
	\cite{ cazorla2019}.	
	
	The similar trends for $i\delta$ = 50 meV suggests that SiC-4H will likely 
	be the stable polytype when $i\delta$ is decreased further for full 
	convergence \cite{Ponce2015}.
	
	We now discuss the EPI correction to the phonon frequencies and energies. 
	As discussed earlier, the correction has two components, adiabatic and 
	non-adiabatic. For insulators and large band gap semiconductors, the 
	non-adiabatic component is negligible and it is justified to consider only 
	the correction from the adiabatic component \cite{Calandra2010adiabatic, 
	Giustino2017}. 

	The adiabatic component is already included in DFPT calculations 
	\cite{Allen2020, Allen2022, Giustino2017, Heid201312}. In the case of 
	SiC-3C, Zywietz et al. \cite{Karch1996} have calculated the vibrational 
	Helmholtz free  energy,  $F_{vib}(T)$, by two different methods, i) 
	generalized bond-charge model (BCM) and ii) DFPT. They show that the 
	 difference in $F_{vib}(T)$ between the two methods is small for SiC-3C. 
	 Indeed, Zywietz et al.  \cite{Karch1996} use this small difference in 
	 $F_{vib}(T)$ for SiC-3C	as the justification to obtain the $F_{vib}(T)$ 
	 for SiC-2H, SiC-4H  and SiC-6H using only BCM. The underlying assumption 
	 being that the  Helmholtz free energy obtained using BCM are likely to be 
	 similar to those obtained using DFPT. The difference in $F_{vib}(T)$ 
	 between SiC-4H and SiC-3C, obtained from the BCM method, is reported to be 
	 \cite{Karch1996} $\Delta  F^{4H-3C}_{vib}(T)$  $\approx$ -0.9 meV/f.u. at 
	 0 K and $\approx$ -3 meV/f.u. at 1200 K \cite{Karch1996}. 

	Recently, Ramakers et al. \cite{Ramakers} 
	have calculated the Helmholtz vibrational free energy differences between 
	SiC polytypes using DFPT. Their values for  $\Delta F^{4H-3C}_{vib}(T)$ 
	are $\approx$ -0.3 meV/f.u. at 0 K and $\approx$ -1.7 meV/f.u. at 1200 K. 
	The DFPT results of Ramakers et al. 	
	\cite{Ramakers} are similar and validate 
	the free energy differences obtained by Zywietz et al. \cite{Karch1996} 
	using BCM. The similar values obtained from BCM and DFPT imply that the former implicitly includes the adiabatic component of the EPI contribution to the phonon energy \cite{Allen2022}.

 	Because the adiabatic component of the EPI contribution to the free energy 
 	differences is very small and the non-adiabatic component of 
	 the  phonon self-energy is negligible in large band-gap semiconductors 
	 \cite{Calandra2010adiabatic, Giustino2017}, we can conclude that the EPI 
	 contribution to the vibrational free energy differences between SiC 
	 polytypes are very small or negligible. 

	With our results, we can assess the importance of the three correction 
	terms, ZPVE, vdW and EPI, to the DFT relative stability order of SiC 
	polytypes. The ZPVE is relatively insensitive to crystal structure 
	\cite{Scalise2019} which is also reflected in the similar Debye 
	temperatures of SiC polytypes \cite{Xu2018, Moruzzi1988}. Its contribution 
	to the energy differences between SiC polytypes is $\sim$ 0-1 meV/f.u. 
	using DFPT and other methods 
	\cite{Ramakers, Karch1996}. The D3(BJ) 
	correction is also relatively insensitive to crystal structure. It 
	contributes $\sim$1-2 mev/f.u. to the energy differences between SiC 
	polytypes. In contrast, EPI correction contributes $\geq$ 22 meV/f.u to the 
	energy differences (Figure \ref{fig2}) indicating much greater sensitivity 
	to crystal structure. Clearly, the EPI term is the most important of the 
	three correction terms to affect the relative stability of SiC 
	polytypes. Hence, EPI contributions must be included in all studies of 
	materials with polymorphs that differ marginally in energy where currently 
	only ZPVE and vdW contributions are included.
	
	After including vdW and EPI corrections to DFT, the relative stability 
	order is SiC-4H, SiC-2H and SiC-3C. The hexagonal SiC-4H is stable over the 
	cubic SiC-3C by $\sim$ 25 meV/f.u. or $\sim$ 2.5 kJ/mol. Hence, by 
	including EPI corrections, the stability of SiC-4H is significantly 
	enhanced when compared to DFT studies, with or without the vdW 
	approximation. 
		
	Our results are consistent with the experimental results of Kleykamp 
	\cite{Kleykamp1998gibbs}. Our results also provide additional 
	motivation to confirm  (or contradict)  Kleykamp's \cite{Kleykamp1998gibbs} 
	results by independent electrochemical	experimental studies. As discussed 
	earlier, electrochemical studies can provide a clear and unambiguous 
	indication of the stable SiC polytype.
	
	It is evident that including EPI contributions using the approximate Eq. 
	\ref{EPI-approx-Abinit} leads to a better match with the experimental 
	results of Kleykamp\cite{Kleykamp1998gibbs}  compared to earlier DFT based 
	studies	\cite{Park1994, Kackell1994, Kawanishi2016, Scalise2019, 
	Heine1991, Rutter1997, Heine1992computational, Ramakers}. Because Eq.  
	\ref{EPI-approx-Abinit} neglects the $1-f_{k+Q}$ factor present in  Eq. 
	\ref{EPI-zero-K-Allen}, our results imply that the neglect of the 
	$1-f_{k+Q}$ factor is not dominant in  determining the differences in EPI 
	contributions to the total energy. This is due to cancellation 
	of errors, especially in structurally similar SiC polytypes where the 
	errors due to the neglect of the $1-f_{k+Q}$ factor are likely to be 
	similar.  However, Eq. \ref{EPI-zero-K-Allen} is the correct expression 
	and must be used once its computation is included in software packages.
	
	Our results have wide applicability. A significant EPI results in ZPR of 
	VBM and other eigenstates in hundreds of meV in several materials 
	\cite{Marini2008,Giustino2010,Antonius2014,Ponce2014b,Ponce2015,Antonius2015,
	Patrick2014,Zacharias2016,Friedrich2015,Allen2018,Tutchton2018,Querales2019,Gonze2020}.
	 It also leads to strong crystral structure dependence of the ZPR(VBM) with 
	 its differences in tens of meV in AlN, BN and GaN polytypes 
	 \cite{Ponce2015, Tutchton2018, Gonze2020}. Because the ZPR(VBM) 
	contributes to  $\Delta E^{ep}_{av}(0)$, it follows that differences in  
	$\Delta E^{ep}_{av}(0)$ (and hence $\Delta E_{EP} (V,0)$) between 
	polymorphs of order of tens of meV/f.u. is a distinct possibility that must 
	be evaluated. Our results for C and SiC polytypes support this suggestion.
	
	It is essential to obtain accurate energy differences between polymorphs to determine their pressure-temperature (P-T) stability regions and phase boundaries. Thus, our results imply that including EPI contributions to the total energy is essential to accurately determine the P-T phase stability of materials in \textit{ab initio} studies.
	
	\section{Conclusion}
		
	We propose and compute a new correction term, due to electron-phonon 
	interactions, to 
	the DFT total energy. We rely on Allen's general formalism that goes beyond 
	the QHA and includes contributions from quasiparticle interactions to the 
	free energy. We show that, for semiconductors and insulators, the 
	EPI contributions to the free energies of electrons and phonons are the 
	corresponding zero-point energy contributions. Using Allen's expression in 
	combination with the Allen-Heine theory for the EPI calculations, we 
	calculate the EPI corrections to the total energy for C, Si and SiC 
	polytypes. The EPI corrections alter the energy differences between 
	polytypes; especially in C and SiC where the EPI strength is significant. 
	Compared to the ZPVE and vdW correction, the EPI correction term is more 
	important in determining the relative stability order of SiC polytypes due 
	to its greater sensitivity to crystal structure. It clearly establishes 
	that in \textit{ab initio} studies SiC-3C is metastable and SiC-4H is the 
	stable polytype, consistent with the experimental results of Kleykamp. Our 
	study enables the inclusion of EPI contribution as a separate term in the 
	free energy expression. This opens the way to go beyond QHA by including 
	the contribution of EPI to all thermodynamic properties.
	
\begin{acknowledgement}

The authors are deeply thankful to Prof. P. B. Allen for sharing a draft of Second Erratum \cite{Allen2022erratum} and for the comments in his named reviews of the earlier versions of the manuscript. We also thank the ``Spacetime'' HPC facilities at IIT Bombay for computational support.

\end{acknowledgement}

\bibliography{references.bib}

\end{document}